\documentclass[aps,prc]{revtex4}
\usepackage{epsfig}
\newcommand{\vsig}{\mbox{\boldmath$\sigma$\unboldmath}}
\newcommand{\veps}{\mbox{\boldmath$\epsilon$\unboldmath}}

\newcommand{\be}{\begin{equation}}
\newcommand{\ee}{\end{equation}}
\newcommand{\bea}{\begin{eqnarray}}
\newcommand{\eea}{\end{eqnarray}}
\newcommand{\bean}{\begin{eqnarray*}}
\newcommand{\eean}{\end{eqnarray*}}

\newcommand{\gapproxeq}{\lower
.7ex\hbox{$\;\stackrel{\textstyle >}{\sim}\;$}}
\newcommand{\lapproxeq}{\lower
.7ex\hbox{$\;\stackrel{\textstyle <}{\sim}\;$}}

\begin{document}

\title{\bf Study of the exotic $\Theta^+$ with polarized photon beams}

\author{Qiang Zhao\footnote{E-mail: Qiang.Zhao@surrey.ac.uk}
}
\affiliation{ Department of Physics, 
University of Surrey, Guildford, GU2 7XH, United Kingdom}

\begin{abstract}

We carry out an analysis of the pentaquark $\Theta^+$
photoproduction with polarized photon beams. 
Kinematical and dynamical aspects are examined 
for the purpose of determining $\Theta^+$'s spin and parity.
It shows that the polarized photon beam asymmetry 
in association with certain dynamical properties of the production
mechanism would provide further information on its quantum numbers. 
Facilities at SPring-8, JLab, ELSA, and ESRF
will have access to them.

\end{abstract}

\maketitle  
\vskip 1.cm


\section{Introduction}

One of the major successes of the constituent quark model 
was its description of the established baryons as states
consisting of three quarks, i.e., 
the minimal number of quarks to form a color-singlet  
``ordinary" baryon. 
Therefore, a baryon 
with strangeness $S=+1$, if it exists, can be regarded as ``exotic". 
The newly-discovered $\Theta^+$~\cite{spring-8,diana,clas,saphir,hicks} 
seems to be the first experimental 
evidence for such a state. In particular, it has 
a quite low mass (1.54 GeV) and a narrow width $< 25$ MeV, which 
certainly brings a strong impact on the conventional
picture for the baryon structures.

Practically, 
the absence of the $S=+1$ baryon in the  three-quark picture 
will not rule out the quark model if an extra
quark pair is present, e.g. $uudd\bar{s}$. 
The possible existence of such an object has been 
discussed before in the literature based on QCD 
phenomenology~\cite{old-papers,gao-ma}.

On the other hand, a baryon with $S=+1$ seems to be a natural 
output of the SU(3) Skyrme model, where the $\Theta^+$
is assigned to be a member of an exotic $\bar{\bf 10}$ multiplet
of spin-parity $1/2^+$ 
in company with the ordinary {\bf 8} and {\bf 10} 
multiplets~\cite{skyrme}.
Remarkably, the quantitative calculation~\cite{dpp} predicts
that the $S=+1$ member has a mass of 1.53 GeV with a narrow width 
of about 9 MeV, which is in good agreement with the experimental
data. But is the $\Theta^+$ the Skyrme model predicted state?

In constrast to the Skyrme model predictions~\cite{skyrme,dpp}, 
the quark model has a rather different 
picture for the exotic $\Theta^+$ due to possible quark flavor-spin 
correlations. For instance, 
if the lowest-mass pentaquark state
is in a relative $S$ wave, 
it will have spin-parity $J^P = \frac{1}{2}^-$. 
Therefore, to accommodate the $\Theta^+$
of $1/2^+$ as predicted by the Skyrme model, asymmetric 
quark correlations seem likely. 
Stancu and Riska~\cite{s-r} have proposed a flavor-spin 
hyperfine interaction between light quarks, which is supposed 
to be strong enough to have the lowest state in the orbital $P$ wave, 
and thus produces the quantum numbers $1/2^+$ for the $\Theta^+$. 
An alternative way to recognize the asymmetric 
quark correlations would be 
via clustered quark structures~\cite{close-tornqvist} 
as suggested by Jaffe and Wilczek~\cite{JW} and Karliner and Lipkin~\cite{k-l}. 
To account for the narrow width of the $\Theta^+$, 
Capstick, Page, and Roberts~\cite{caps} have proposed 
that width suppression in $\Theta^+\to K N$ would be due to the isospin symmetry 
violating if the $\Theta^+$ is an isotensor state. 
However, the experimental analysis from SAPHIR~\cite{saphir} does not 
find signals for $\Theta^{++}$ in $\gamma p\to \Theta^{++} K^-$. 
It thus presents a challenge for such an isopsin-symmetry-violating solution.
Nevertheless, even smaller widths for the $\Theta^+$ are 
claimed in the reports of Refs.~\cite{nussinov,gothe,asw,hk}.
Carlson {\it et al.}~\cite{carlson} 
have shown the anti-decuplet $\bar{\bf 10}$ with $J^P=\frac{1}{2}^-$
can lead to the same mass ordering $\Theta\to N\to \Sigma\to \Xi$,
as in the Skyrme model. However, the question about the $\Theta^+$'s
narrow width has not been answered.

Apart from those efforts mentioned above, 
a large number of theoretical attempts have recently been devoted to 
understanding the nature of the $\Theta^+$.  
Perspectives of the pentaquark properties and their consequences
have been discussed in a series of activities
in the Skyrme model~\cite{bfk,praszalowicz,cohen,ikor},  
quark models~\cite{cheung,glozman,jm,hzyz,okl,dp,bijker}, 
and other phenomenologies~\cite{kk,gk},
which are essentially embedded on
the phenomenological assumptions for the $\Theta^+$'s quantum numbers,
either to be a Skyrme $\bar{\bf 10}$ state of $1/2^+$ 
or pentaquark of $1/2^-$, $1/2^+$, $3/2^-$, $3/2^+, \cdots$.
QCD sum rule studies~\cite{zhu,mnnrl,sdo} 
and lattice QCD calculations~\cite{sasaki,cfkk} are also reported. 
It is interesting to note that both QCD sum rules and lattice calculations
are in favor of the $\Theta^+$ to be $1/2^-$. 
Note that the spin and parity of the $\Theta^+$ have not yet been 
unambiguously determined in experiment. 
This means that it is difficult to rule out any of those possibilities.
More essentially, such a situation also raises questions such as 
whether, or under what circumstances, there is any correspondence 
between the Skyrme and quark pictures. 
Or if the $\Theta^+$ is a pentaquark of $1/2^+$, since there must be angular
momentum to overcome the intrinsic parity of $uudd\bar{s}$ 
it implies there must be $3/2^+$ partners. Then 
the questions could be how big the spin-orbit mass splitting 
between $1/2^+$ and $3/2^+$ would be, and which state was 
measured here~\cite{close1,jm}.

This situation eventually suggests that
an explicit experimental confirmations 
of the quantum numbers of the $\Theta^+$ are not only essentially important for 
establishing the status of $\Theta^+$ on a fundamental basis~\cite{asw,aasw}, 
but also important for any progress in understanding 
its nature, and the existence of its other 
partners~\cite{na49}. 
A number of efforts to study the $\Theta^+$ in meson photoproduction and
meson-nucleon scattering were  made recently 
for such a purpose~\cite{pr,hyodo,nam,liu-ko,oh}, 
where the $\Theta^+$ production in photonuclear reactions 
or $\pi$-$N$ scattering were estimated. 
The model calculations of the cross sections suggest
a significant difference between $1/2^+$ and $1/2^-$, which would certainly 
make sense for the attempt at determining the quantum numbers of $\Theta^+$.
This could be the first reference for distinguishing these two 
configurations, but should be taken with caution.
The reason is that little knowledge about the $\Theta^+$
form factor is available and its total width still has large uncertainties. 
Nevertheless, the role played by other mechanism, e.g., $K^*$ exchange,
is almost unknown, to which, however, the cross section is very sensitive. 
Due to such complexities, other observables should be measured, which 
will on the one hand provide more information for 
the $\Theta^+$'s quantum numbers, 
and on the other hand provide constraints on theoretical phenomenologies.

In this work, we will study the photoproduction of  
$\Theta^+$ with polarized photon beams, and 
look for further information about the $\Theta^+$ which 
would be useful for the determination of its quantum numbers.
First, in Sec. II we will examine the photon polarization transfer by studying 
the $\Theta^+$ decay in terms of its density matrix elements.
In Sec. III, we will discuss the reaction mechanisms and 
analyze the polarized beam asymmetries 
in association with the differential cross section, 
for which different quantum numbers for $\Theta^+$ will lead to 
different features. 
Discussions and summaries will be given in Sec. IV.

\section{Density matrix elements for $\Theta^+$ decay}

The $\Theta^+$ is found to have a narrow width $<25$ MeV, which 
in principle makes the experimental access to its kinematic 
reconstruction easier than a broad state.
In this sense, the production of $\Theta^+$ can be treated as a
two-body reaction $\gamma n\to K^-\Theta^+$. 
This also gives access to the $\Theta^+$ production 
with polarized photon beams. 
Experimental efforts
can be based on the present electromagnetic probes available 
at most of those experimental facilities, e.g., SPring-8, JLab, 
ELSA, and ESRF with their full-angle detectors. 
To be closely related to the experimental measurements and analyses, 
we will start with the experimentally measurable density matrix elements, 
which is useful for separating the kinematical information 
from the dynamical one, and look for signals which can help to pin down 
the quantum numbers of $\Theta^+$.

Defining the $z$-axis as the photon momentum direction, 
and the reaction plane in $x$-$z$ in the c.m. system of $\gamma$-$n$, 
the transition amplitude for $\gamma n\to K^-\Theta^+$
can be expressed as 
\be
\label{heli-amp}
T_{\lambda_\theta,\lambda_\gamma \lambda_N}\equiv
\langle \Theta^+, \lambda_\theta, {\bf P}_\theta; K^-, \lambda_0, {\bf q} 
| \hat{T} | n, \lambda_N, {\bf P}_i; \gamma, \lambda_\gamma, {\bf k}\rangle ,
\ee
where $\lambda_\gamma=\pm 1$, $\lambda_N\pm 1/2$, 
$\lambda_0=0$, and $\lambda_\theta$ 
are helicities 
of photon, neutron, $K^-$, and $\Theta^+$, respectively.

The decay of $\Theta^+$ into $K^+ n$ is analyzed in the helicity frame, 
i.e. in the $\Theta^+$ rest frame;
the azimuthal angles $\theta$ and $\phi$ of $K^+$ are 
defined with respect to the $\Theta^+$ three momentum ${\bf P}_\theta$ 
in the $\gamma$-$n$ c.m. frame. 
The decay of $\Theta^+$ contains information for its spin and parity. 
The strategy here is to analyze three possible configurations
for $\Theta^+$, and compare their consequence 
in the $K^+ n$ angular distributions with polarized photon beams. 

The decay matrix can be written as
$ 
\langle n, s_f, {\bf P}_\theta-{\bf p}^\prime| \hat{t} 
|\Theta^+, s_\theta, {\bf P}_\theta\rangle, 
$
where $\hat{t}$ denotes the strong transition operator.
For the $\Theta^+$ of $1/2^-$, $1/2^+$, and $3/2^+$, we apply
the following effective Lagrangians to 
derive the transition operators:
\bea
\label{lagrangian}
{\cal L}_{eff}(1/2^-) &=& g_{\Theta NK} \bar{\Theta} N K + \mbox{H.c.},
\nonumber\\
{\cal L}_{eff}(1/2^+) &=& g_{\Theta NK} 
\bar{\Theta} \gamma_\mu\gamma_5 \partial^\mu K N  + \mbox{H.c.},
\nonumber\\
{\cal L}_{eff}(3/2^+) &=& g_{\Theta NK} 
\bar{\Theta}^\mu (4g_{\mu\nu}-\gamma_\mu\gamma_\nu)
\partial^\nu K N  + \mbox{H.c.},
\eea
where $\bar{\Theta}$, $N$ and $K$ denote the field of $\Theta^+$, neutron
and $K^-$, respectively. 
The nonrelativistic expansion in the $\Theta^+$ rest frame 
(${\bf P}_\theta=0$) gives:
\bea
\hat{t}^{(s,1/2^-)}&=& {\cal C} , \nonumber\\
\hat{t}^{(p,1/2^+)}&=& {\cal C} \vsig\cdot {\bf p}^\prime , \nonumber\\
\hat{t}^{(p,3/2^+)}&=& {\cal C} {\bf S}\cdot {\bf p}^\prime ,
\eea
where ${\cal C}$ is a spin and angular
independent factor, and ${\bf p}^\prime$ is the three momentum carried by 
$K^+$ in the $\Theta^+$ rest frame.
Note that ${\cal L}_{eff}(1/2^-)$ is different from the treatment 
of Ref.~\cite{nam}, where the effective Lagrangian is obtained 
by removing the $\gamma_5$ matrix from the pseudovector/pseudoscalar
coupling, though the $S$-wave decay leads to the same form for the 
transition operator.

\subsection{$\Theta^+$ of $1/2^+$}

Starting with the $1/2^+$ configuration, 
the decay transition can be expressed as
\bea 
\label{decay-1}
\hat{R}_{s_f, s_\theta} &\equiv &
\langle n, s_f, {\bf P}_\theta-{\bf p}^\prime| \hat{t}^{(p,1/2^+)} 
|\Theta^+, s_\theta, {\bf P}_\theta=0\rangle
\nonumber\\
&=& {\cal C} [ 2 s_\theta D^{1*}_{00}(\phi,\theta,-\phi) 
\delta_{s_f,s_\theta}
-\sqrt{2} D^{1*}_{10}(\phi,\theta,-\phi)\delta_{s_f,s_\theta-1}
+\sqrt{2} D^{1*}_{-10}(\phi,\theta,-\phi)\delta_{s_f,s_\theta+1}] ,
\eea
where we use the short-hand
symbol $\hat{R}_{s_f, s_\theta}$ to denote the transition elements and 
$D^1_{MN}(\phi,\theta,-\phi)$ is the Wigner rotation function
with the convention of Ref.~\cite{rose}. Namely, 
 $D^I_{MN}(\alpha,\beta,\gamma)=d^I_{MN}(\beta) e^{-i(M\alpha+N\gamma)}$.
Under this convention, those rotation functions will be
\bea
D^1_{00}(\phi,\theta,-\phi) &=& d^1_{00}(\theta)=\cos\theta \ ,\nonumber\\
D^1_{10}(\phi,\theta,-\phi) &=& d^1_{10}(\theta)e^{-i\phi}
=-\frac{\sin\theta}{\sqrt{2}}e^{-i\phi} \ , \nonumber\\
D^1_{-10}(\phi,\theta,-\phi) &=& d^1_{-10}(\theta)e^{i\phi}
=\frac{\sin\theta}{\sqrt{2}}e^{i\phi} .
\eea

The angular distribution of $\Theta^+\to K^+ n$
can be then described by the density matrix elements 
$\rho_{s_\theta,s^\prime_\theta}(\Theta^+)$:
\bea
W(\theta,\phi) &=& 
\sum_{s_f, s_f^\prime; s_\theta, s_\theta^\prime}
\hat{R}_{s_f,s_\theta} \rho_{s_\theta,s^\prime_\theta}(\Theta^+)
\hat{R}^*_{s_f^\prime,s_\theta^\prime} \nonumber\\
&=& \sum_{s_\theta, s_\theta^\prime}
\big\{ \hat{R}_{-\frac 12,s_\theta} \rho_{s_\theta,s^\prime_\theta}
\hat{R}^*_{-\frac 12,s_\theta^\prime}
+ \hat{R}_{\frac 12,s_\theta} \rho_{s_\theta,s^\prime_\theta}
\hat{R}^*_{-\frac 12,s_\theta^\prime} \nonumber\\
&& + \hat{R}_{\frac 12,s_\theta} \rho_{s_\theta,s^\prime_\theta}
\hat{R}^*_{\frac 12,s_\theta^\prime}
+ \hat{R}_{-\frac 12,s_\theta} \rho_{s_\theta,s^\prime_\theta}
\hat{R}^*_{\frac 12,s_\theta^\prime} \big\} .
\eea
Substituting Eq.~(\ref{decay-1}) into the above, 
and after some simple algebra, we arrive at 
\bea
\label{distr-1}
W(\theta,\phi) &=& 
[1-\sin 2\theta\cos 2\phi]\rho_{-\frac 12,-\frac 12}
+[1+\sin 2\theta\cos 2\phi] \rho_{\frac 12,\frac 12} \nonumber\\
&& -[\cos^2\theta - \sin 2\theta e^{-2i\phi}]
\rho_{-\frac 12, \frac 12} 
-[\cos^2\theta -\sin^2\theta e^{2i\phi}]
\rho_{\frac 12, -\frac 12} ,
\eea
where to be consistent with the conventions for the transition amplitudes 
defined by Eq.~(\ref{heli-amp}), we have labeled the density matrix elements
by the helicities of the particles. 
Since the helicity direction is defined as the momentum of the $K^-$
in the final state, 
the only change is $s_\theta\to -\lambda_\theta$ and
$s_\theta^\prime\to -\lambda_\theta^\prime$. 

The density matrix elements for the $\Theta^+$ can be related to 
the photon density matrix elements in the $\Theta^+$ production:
\be
\rho_{\lambda_\theta,\lambda_\theta^\prime}
=\frac{1}{N}\sum_{\lambda_N,\lambda_\gamma,\lambda_\gamma^\prime}
T_{\lambda_\theta,\lambda_\gamma \lambda_N} 
\rho_{\lambda_\gamma\lambda_\gamma^\prime}(\gamma)
T^*_{\lambda_\theta^\prime,\lambda_\gamma^\prime \lambda_N} \ ,
\ee
where $N=\frac 12 \sum_{\lambda_\theta,\lambda_\gamma,\lambda_N}
|T_{\lambda_\theta,\lambda_\gamma \lambda_N}|^2$ is the normalization
factor, and it has conventionally a factor 1/2 difference from 
the initial spin average and final spin summation~\cite{schilling}.
Note that 
$\rho(\gamma)=\frac 12({\bf 1} +{\bf P}_\gamma\cdot\vsig)$
is determined by the polarization direction ${\bf P}_\gamma$
and the eigenvalue of the photon polarization states. 
As discussed in Ref.~\cite{schilling}, 
for circularly polarized photons, ${\bf P}_\gamma=P_\gamma (0,0,\pm 1)$, 
while for linearly polarized photons, 
${\bf P}_\gamma=P_\gamma (-\cos 2\Phi, -\sin 2\Phi, 0)$,
where $\Phi$ is the angle between the photon polarization vector
$\veps_\gamma$ and the $\Theta^+$ production plane
($x$-$z$ plane)~\cite{schilling}, 
and $P_\gamma$ denotes the photon polarization degree in experiment.
Given the eigenvalues for different polarization components of the photons, 
the corresponding  density matrix elements for the $\Theta^+$ 
can be defined:
\bea
\rho^0_{\lambda_\theta,\lambda_\theta^\prime}
&=& \frac{1}{2N}\sum_{\lambda_\gamma\lambda_N}
T_{\lambda_\theta,\lambda_\gamma \lambda_N} 
T^*_{\lambda_\theta^\prime,\lambda_\gamma \lambda_N} , \\
\rho^1_{\lambda_\theta,\lambda_\theta^\prime}
&=& \frac{1}{2N}\sum_{\lambda_\gamma\lambda_N}
T_{\lambda_\theta,-\lambda_\gamma \lambda_N} 
T^*_{\lambda_\theta^\prime,\lambda_\gamma \lambda_N} , \\
\rho^2_{\lambda_\theta,\lambda_\theta^\prime}
&=& \frac{i}{2N}\sum_{\lambda_\gamma\lambda_N}
\lambda_\gamma T_{\lambda_\theta,-\lambda_\gamma \lambda_N} 
T^*_{\lambda_\theta^\prime,\lambda_\gamma \lambda_N} , \\
\rho^3_{\lambda_\theta,\lambda_\theta^\prime}
&=& \frac{1}{2N}\sum_{\lambda_\gamma\lambda_N}
\lambda_\gamma T_{\lambda_\theta,\lambda_\gamma \lambda_N} 
T^*_{\lambda_\theta^\prime,\lambda_\gamma \lambda_N} .
\eea
Together with the specified polarization direction ${\bf P}_\gamma$, 
the $\Theta^+$ decay distribution for the linearly polarized photon
beams can be expressed as
\be
\label{distr-3}
W(\theta,\phi) = W^0(\cos\theta, \phi, \rho^0)
-P_\gamma\cos 2\Phi W^1(\cos\theta, \phi, \rho^1)
-P_\gamma\sin 2\Phi W^2(\cos\theta, \phi, \rho^2) ,
\ee
where $W^\alpha$ ($\alpha=0$, 1, 2, 3) corresponds to 
the decay distributions with different photon polarizations.
Easily, it can be seen that $W^0$ 
is for the distribution with unpolarized photon, 
$W^{1,2}$ with linearly polarized photon, and $W^3$ with circularly
polarized photon.
The explicit expressions for $W^\alpha$ is given by 
Eq.~(\ref{distr-1}), and can be simplified 
by taking into account parity conservation:
\be
\label{parity-cons}
T_{\lambda_\theta,\lambda_\gamma \lambda_N}
=(-1)^{-\lambda_\gamma-(\lambda_\theta -\lambda_N)}
T_{-\lambda_\theta,-\lambda_\gamma -\lambda_N}  \ ,
\ee
which leads to 
\be
\rho^\alpha_{-\frac 12, -\frac 12} = \rho^\alpha_{\frac 12, \frac 12} \ , \ \ \ \
\rho^\alpha_{-\frac 12, \frac 12}=-\rho^\alpha_{\frac 12, -\frac 12} \ , 
\ee
for $\alpha=0, 1$; 
and 
\be
\rho^\alpha_{-\frac 12, -\frac 12} = -\rho^\alpha_{\frac 12, \frac 12} \ , \ \ \ \
\rho^\alpha_{-\frac 12, \frac 12}=\rho^\alpha_{\frac 12, -\frac 12} \ , 
\ee
for $\alpha=2, 3$.
On the other hand, the density matrix elements
must be Hermitian:
\be
\rho^\alpha_{\lambda_\theta,\lambda_\theta^\prime} \equiv
\rho^{\alpha *}_{\lambda_\theta^\prime,\lambda_\theta} \ ,
\ee
which leads to:
\be 
\mbox{Re}\rho^\alpha_{-\frac 12, \frac 12}
=\mbox{Re}\rho^\alpha_{\frac 12, -\frac 12}\equiv 0 , \ \ \ \ \
\mbox{Im}\rho^\alpha_{-\frac 12, \frac 12}
=-\mbox{Im}\rho^\alpha_{\frac 12, -\frac 12}, \ \ \ \ \ \mbox{for} \ \ \alpha=0,1
\ee
and
\be 
\mbox{Im}\rho^\alpha_{-\frac 12, \frac 12}
=\mbox{Im}\rho^\alpha_{\frac 12, -\frac 12}\equiv 0 , \ \ \ \ \
\mbox{Re}\rho^\alpha_{-\frac 12, \frac 12}
=\mbox{Re}\rho^\alpha_{\frac 12, -\frac 12}, \ \ \ \ \ \mbox{for} \ \ \alpha=2, 3. 
\ee
Also note that elements with $\lambda_\theta=\lambda_{\theta^\prime}$ are always
real. 
Thus, 
the angular distribution for $\Theta^+$ decay 
with different polarizations will be:
\bea
\label{distr-p-wave1}
W^0(\cos\theta, \phi, \rho^0) &=& 
2\rho^0_{\frac 12,\frac 12}
+2\sin^2\theta\sin 2\phi
\mbox{Im}\rho^0_{\frac 12, -\frac 12}  \ , \nonumber\\
W^1(\cos\theta, \phi, \rho^1) &=& 
2\rho^1_{\frac 12,\frac 12}
+2\sin^2\theta\sin 2\phi
\mbox{Im}\rho^1_{\frac 12, -\frac 12}  \ , \nonumber\\
W^2(\cos\theta, \phi, \rho^2) &=& 
-2\sin 2\theta \cos 2\phi \rho^2_{\frac 12,\frac 12}
-2[\cos^2\theta + \sin^2\theta\cos 2\phi]
\mbox{Re}\rho^2_{\frac 12, -\frac 12} \ , \nonumber\\
W^3(\cos\theta, \phi, \rho^3) &=& 
-2[\cos^2\theta + \sin^2\theta\cos 2\phi]
\mbox{Re}\rho^3_{\frac 12, -\frac 12} \ , 
\eea
where $W^3$ does not contribute to the linearly polarized photon
reaction. In the above equation, we have used the relation 
$\rho^3_{-\frac 12,-\frac 12}=\rho^3_{\frac 12,\frac 12}\equiv 0$.

\subsection{$\Theta^+$ of $1/2^-$}

Assuming that the $\Theta^+$ has a spin-parity $1/2^-$, 
the decay transition will not involve spin operators, 
and can be simply expressed as:
\be
\hat{R}_{s_f, s_\theta} \equiv 
\langle n, s_f, {\bf P}_\theta-{\bf p}^\prime| \hat{t}^{(s,1/2^-)} 
|\Theta^+, s_\theta, {\bf P}_\theta=0 \rangle
= {\cal C} \delta_{s_f,s_\theta} .
\ee
Following the above procedure, 
we have the angular distribution of $\Theta^+$ decay:
\bea
\label{distr-s-wave}
W^0(\cos\theta, \phi, \rho^0) &=& 
2 \rho^0_{\frac 12,\frac 12} \ , \nonumber\\
W^1(\cos\theta, \phi, \rho^1) &=& 
2 \rho^1_{\frac 12,\frac 12} \ , \nonumber\\
W^2(\cos\theta, \phi, \rho^2) &=& 
2 \mbox{Re}\rho^2_{\frac 12, -\frac 12} \ , \nonumber\\
W^3(\cos\theta, \phi, \rho^3) &=& 
2 \mbox{Re}\rho^3_{\frac 12, -\frac 12} \ , 
\eea
where the isotropic distribution of the $\Theta^+$ decay is 
due to the relative $S$-wave between the $K^+$ and $n$.

\subsection{$\Theta^+$ of $3/2^+$}

For $\Theta^+$ of spin-parity $3/2^+$, 
its decay transition operator is given by the Rarita-Schwinger
field transition to a spinor and a vector field:
\bea
\hat{R}_{s_f, s_\theta} &\equiv &
\langle n, s_f, {\bf P}_\theta-{\bf p}^\prime
| {\bf S}\cdot{\bf p}^\prime |\Theta^+, s_\theta, {\bf P}_\theta=0 \rangle
\nonumber\\
&=& {\cal C} \sum_\lambda \langle 1 \lambda, \frac 12 s_f | \frac 32 s_i\rangle
{\bf e}_\lambda\cdot {\bf p}^\prime \nonumber\\
&=& {\cal C}^\prime 
\sum_\lambda \langle 1 \lambda, \frac 12 s_f | \frac 32 s_i\rangle
D^{1*}_{\lambda 0}(\phi,\theta,-\phi),
\eea
where ${\bf e}_\lambda$ is the spherical vector presentation of 
an orthogonal set of cartesian unit vectors $\hat{x}$, 
$\hat{y}$, and $\hat{z}$:
\be
{\bf e}_{\pm 1}=\mp \frac{1}{\sqrt 2}(\hat{x}\pm \hat{y}), \ \ \ 
{\bf e}_0=\hat{z} .
\ee
Following the same strategy as the above, 
the $\Theta^+$ decay distribution can be obtained:
\bea
\label{p-wave2a}
W^\alpha(\cos\theta, \phi, \rho^\alpha) &=& 
\left(\frac 13 +\cos^2\theta\right)\rho^\alpha_{\frac 12,\frac 12}
+\sin^2\theta \rho^\alpha_{\frac 32,\frac 32}
-\frac 13\sin^2\theta \sin 2\phi\mbox{Im}\rho^\alpha_{\frac 12,-\frac 12}
-\sin^2\theta \sin 2\phi
\mbox{Im}\rho^\alpha_{\frac 32,-\frac 32} \nonumber\\
&& +\frac{2}{\sqrt 3}\sin 2\theta\cos\phi 
\mbox{Re}\rho^\alpha_{\frac 12,\frac 32}
-\frac{2}{\sqrt 3}\sin^2\theta\cos 2\phi
\mbox{Re}\rho^\alpha_{\frac 12,-\frac 32}
-\frac{2}{\sqrt 3}\sin 2\theta\sin\phi
\mbox{Im}\rho^\alpha_{\frac 12,-\frac 32}
\eea
for $\alpha=0,1$, and 
\bea
\label{p-wave2b}
W^\alpha(\cos\theta, \phi, \rho^\alpha) &=& 
\frac 13 \sin 2\theta \cos \phi \rho^\alpha_{\frac 12,\frac 12}
+\left(\frac 43\cos^2\theta -\frac 13 \sin^2\theta\cos 2\phi\right)
\mbox{Re}\rho^\alpha_{\frac 12,-\frac 12}
-\sin^2\theta\cos 2\phi \mbox{Re}\rho^\alpha_{\frac 32,-\frac 32}
\nonumber\\
&& + \frac{2}{\sqrt 3}\sin^2\theta \mbox{Re}\rho^\alpha_{\frac 12,\frac 32}
-\frac{2}{\sqrt 3}\sin 2\theta\sin\phi \mbox{Im}\rho^\alpha_{\frac 12,\frac 32}
\nonumber\\
&& - \frac{2}{\sqrt 3}\sin 2\theta\cos\phi 
\mbox{Re}\rho^\alpha_{\frac 12,-\frac 32}
-\frac{2}{\sqrt 3}\sin^2\theta\sin 2\phi 
\mbox{Im}\rho^\alpha_{\frac 12,-\frac 32} 
\eea
for $\alpha=2,3$, and with $\rho^3_{\frac 12,\frac 12}\equiv 0$.
In the above two equations, parity conservation relation and 
the requirement of $\rho^\alpha$ to be Hermitian have been used, 
and the elements are presented in the helicity frame.
We note that for elements 
$\rho^\alpha_{\pm\lambda_\theta,\pm\lambda_\theta^\prime}$
with $|\lambda_\theta|\neq |\lambda_\theta^\prime$, their 
real (imaginary) parts are not independent of each other
due to parity conservation and Hermitian.
Also, it is found that 
$\mbox{Im}\rho^{0,1}_{-\frac 12,-\frac 32}=\mbox{Im}\rho^{0,1}_{\frac 12,\frac 32}$ 
is not necessarily zero. However, this term 
is canceled out, which could suggest that there is no experimental 
access to this quantity by using the polarized photon beams.

\section{Polarization observables}

The advantage of polarization measurement is that 
the interference between different transition amplitudes
can be highlighted by the asymmetries. 
Therefore, additional information on the transition dynamics can be 
gained. In theory, the study of the polarized beam asymmetries is also 
useful to partially avoid uncertainties arising from the form factors, 
while the availability of experimental data
will provide constraints on parameters introduced in a phenomenology.

With the polarized photon beams
the polarized beam asymmetries can be measured 
via Eqs.~(\ref{distr-p-wave1}),~(\ref{distr-s-wave}),~(\ref{p-wave2a}) and 
~(\ref{p-wave2b}). 
In particular,  with the linearly polarized photons, 
$W^{0,1,2}$ can be measured. 

In principle, one can choose the polarization direction of the 
photons to derive as much information as possible. 
Equation~\ref{distr-3} is general for linearly polarized photons.
We thus can choose $\Phi=90^\circ$ to polarize the photons
along the $y$-axis, and $\Phi=0^\circ$ to polarize
the photons along the $x$-axis. 
By integrating
over $\theta$ and $\phi$, namely, summing over all the experimental events, 
we have respectively:
\bea
\bar{W}_\perp(\Phi=90^\circ,\rho) &=& 
\int_{\theta=0}^\pi\int_{\phi=0}^{2\pi} d\Omega W^0(\cos\theta, \phi, \rho^0)
-P_\gamma \int_{\theta=0}^\pi\int_{\phi=0}^{2\pi} d\Omega W^1(\cos\theta, \phi, \rho^1) 
\nonumber\\
&=&\bar{W}^0(\rho^0)+P_\gamma \bar{W}^1(\rho^1) \ ,
\eea
and 
\be
\bar{W}_\parallel(\Phi=0^\circ,\rho) 
=\bar{W}^0(\rho^0)-P_\gamma \bar{W}^1(\rho^1) \ ,
\ee
which are cross sections for the two polarizations.
In experiment, the polarized beam asymmetry can be defined as
\be
\Sigma_A\equiv 
\frac{\bar{W}_\perp(\Phi=90^\circ,\rho)-\bar{W}_\parallel(\Phi=0^\circ,\rho)}
{\bar{W}_\perp(\Phi=90^\circ,\rho)+\bar{W}_\parallel(\Phi=0^\circ,\rho)}
=\frac{\bar{W}^1(\rho^1)}{\bar{W}^0(\rho^0)} \ ,
\ee
where $(\bar{W}_\perp+\bar{W}_\parallel )$ 
corresponds to the unpolarized cross section.

For the circumstance where $\Theta^+$ has spin-parity $\frac 12^-$ and 
$\frac 12^+$, the expression for the polarized beam asymmetry is the same:
\be
\label{asymm-1}
\Sigma_A=\frac{\rho^1_{\frac 12,\frac 12}}{\rho^0_{\frac 12,\frac 12}} \ .
\ee
However, this does not suggest that these two configurations 
will have the same asymmetries. The values for the elements
will be determined by underlying dynamics. 
Also note that our convention of polarized beam asymmetry 
has a sign difference as that of Ref.~\cite{walker}. It can be seen 
by the decomposition of $\rho^1_{\frac 12,\frac 12}$
in terms of the transition amplitudes $T$:
$\rho^1_{\frac 12,\frac 12}=\frac{1}{N}\mbox{Re}
\{-T_{-\frac 12, 1-\frac 12}T^*_{\frac 12, 1\frac 12} 
+T_{-\frac 12, 1\frac 12}T^*_{\frac 12,1-\frac 12}\}
=-\Sigma_W\rho^0_{\frac 12,\frac 12}$. 
The expression 
is essentially the same as that derived for pseudoscalar meson photoproduction
on the nucleon.
This is understandable since the spins of all the particles
are the same as, e.g., $\gamma n\to \pi^- p$. 
Equation~\ref{asymm-1} also implies a direct access to 
element $\rho^1_{\frac 12,\frac 12}$, which is normalized by 
the differential cross section.

For $3/2^+$, the polarized beam asymmetry is 
\be
\Sigma_A(\frac 32^+)=\frac{\rho^1_{\frac 12,\frac 12}+\rho^1_{\frac 32,\frac 32}}
{\rho^0_{\frac 12,\frac 12}+\rho^0_{\frac 32,\frac 32}} \ ,
\ee
where $2(\rho^0_{\frac 12,\frac 12}+\rho^0_{\frac 32,\frac 32})=1$ 
is the normalized cross section.

For  $3/2^+$, explicit angular dependence is introduced into 
the $\Theta^+$ decay distribution by 
the photon polarization transfer, which makes it much more easier 
to clarify the configuration in experiment.
In particular, for the case that $\rho^0_{\frac 32, \frac 32}$
is much smaller than $\rho^0_{\frac 12, \frac 12}$, 
a clear signal can be seen by the $\cos^2\theta$ distribution
with the $\phi$-angle distribution integrated out at 
$\theta_{c.m.}=0^\circ$:
\be
\label{parity-3}
W(\theta)=2\pi \Big\{ (\frac 13 \rho^0_{\frac 12, \frac 12}
+\rho^0_{\frac 32, \frac 32}) +
\cos^2 \theta (\rho^0_{\frac 12, \frac 12}
- \rho^0_{\frac 32, \frac 32})\Big\}.
\ee
If $\rho^0_{\frac 12, \frac 12}$ and 
$\rho^0_{\frac 32, \frac 32}$ are comparable, 
ambiguities will arise, and dynamic aspects have to be taken into 
account along with the analyses based on kinematics.

Elements $\rho^2$ can be also measured by polarizing the photon
along $\Phi=\pm 45^\circ$. For these three configurations, 
the polarization asymmetry can be defined as
\be
\Sigma_B(\frac 12^-)
=\frac{\mbox{Re}\rho^2_{\frac 12,-\frac 12}}
{\rho^0_{\frac 12,\frac 12}} \ ,
\ee
\be 
\Sigma_B(\frac 12^+)
=\frac{-\mbox{Re}\rho^2_{\frac 12,-\frac 12}}
{3\rho^0_{\frac 12,\frac 12}} \ ,
\ee
and
\be
\Sigma_B(\frac 32^+)
=\frac{2\mbox{Re}\rho^2_{\frac 12,-\frac 12}
+2\sqrt{3}\mbox{Re}\rho^2_{\frac 12,\frac 32}}
{3(\rho^0_{\frac 12,\frac 12}+\rho^0_{\frac 32,\frac 32})} \,
\ee
where different elements can be detected.

For $\Theta^+$ of $3/2^+$, it is likely that 
experimental analyses of its decay distributions has been informative 
of its properties. 
Therefore, as follows, we will concentrate on the production of 
the $1/2^+$ and $1/2^-$ configurations for which 
the transfer of the beam polarized to the $\Theta^+$ 
can only be distinguished through their different dynamical properties.
Also, we will adopt the 
convention of Ref.~\cite{walker} to include the additional sign
for the polarized beam asymmetry.

\subsection{Polarized beam asymmetry for $\Theta^+$ of $1/2^+$ in the Born limit}

\begin{figure}
\begin{center}
\epsfig{file=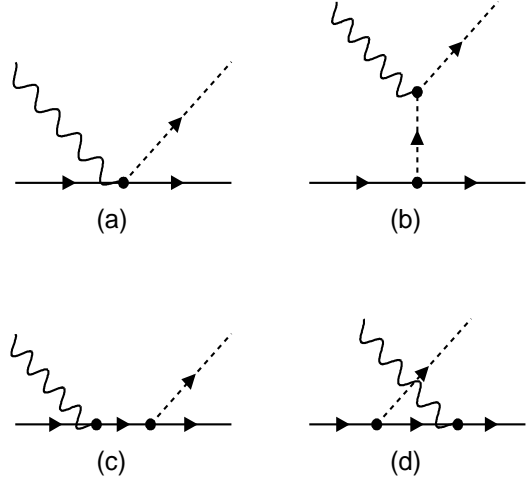, width=10cm,height=9.cm}
\caption{Feynman diagrams for $\Theta^+$ photoproduction in the Born 
approximation. 
}
\protect\label{fig:(1)}
\end{center}
\end{figure}

For the $\Theta^+$ of $1/2^+$, the effective Lagrangian introduces
four transition amplitudes in the Born approximation limit
as shown by Fig.~\ref{fig:(1)}. The leading terms are
 similar to the case of $\gamma n\to \pi^- p$, which would be useful 
for the analyses.  
The transition amplitudes can be expressed as:
\be
{\cal M}_{fi}=M^c + M^t + M^s + M^u  \ ,
\ee
where the four transitions are given by
\bea
M^c & =& ie_0g_{\Theta NK} \bar{\Theta}\gamma_\mu\gamma_5 A^\mu N K ,\nonumber\\
M^t & =&  \frac{ie_0g_{\Theta NK}}{t-M_K^2}\bar{\Theta}
\gamma_\mu\gamma_5 (q-k)^\mu (2q-k)_\nu A^\nu N K,\nonumber\\
M^s & =&  -g_{\Theta NK}\bar{\Theta} \gamma_\mu\gamma_5\partial^\mu K
\frac{[\gamma\cdot(k + P_i) +M_n]}{s-M_n^2} 
\left[ e_n\gamma_\alpha + \frac{i{\kappa}_n }{2M_n}\sigma_{\alpha\beta} 
k^\beta\right] A^\alpha N, \nonumber\\
M^u & =&  -g_{\Theta NK}\bar{\Theta} 
\left[ e_\theta\gamma_\alpha + \frac{i{\kappa}_\theta }{2M_\Theta}
\sigma_{\alpha\beta} k^\beta\right] A^\alpha 
\frac{[\gamma\cdot(P_f-k)+M_\Theta]}{u-M_\Theta^2} 
\gamma_\mu\gamma_5\partial^\mu K N,
\eea
where $e_0$ is the positive unit charge. In the {\it s}-channel
the vector coupling vanishes since $e_n=0$. We define
the coupling constant $g_{\Theta NK}\equiv g_A M_n/f_\theta$ with 
the axial vector coupling $g_A=5/3$, while the decay constant is given by: 
\be
f_\theta=g_A\left(1-\frac{p_0}{E_n+M_n}\right) 
\left[ \frac{|{\bf p}^\prime|^3(E_n+M_n)}
{4\pi M_\Theta\Gamma_{\Theta^+\to K^+ n}}\right]^{1/2},
\ee
where ${\bf p}^\prime$ and $p_0$ are momentum and energy of the kaon
in the $\Theta^+$ rest frame, and $E_n$ is the energy of the neutron.
For the range of $\Gamma_{\Theta^+\to K^+ n}=5$ to 25 MeV, 
$g_{\Theta NK}=2.09$ to 4.68. We adopt $g_{\Theta NK}=2.96$ for 
$\Gamma_{\Theta^+\to K^+ n}=10$ MeV in the calculations.

The $\Theta^+$'s magnetic moment 
 $\mu_\theta$ [$=(1+\kappa_\theta)/2M_\Theta$]
is estimated in the model of Jaffe and Wilczek~\cite{JW}:
\be
\mu_\theta\vsig\equiv 
\sum_{i=1}^3 \frac{e_i}{2m_i}(\vsig_i +{\bf l}_i),
\ee
where $e_1$ ($e_2$) and $m_1$ ($m_2$) denote the charge and 
mass of the $(ud)$ clusters, and $e_3$ and $m_3$ of the $\bar{s}$; 
$\vsig=2 {\bf s}$ denotes the Pauli matrices. 
For $(ud)(ud)\bar{s}$, the isospin wavefunction 
belongs to SU(3) symmetric representation 
$\bar{\bf 10}$, and $u$-$d$ couples to spin zero clusters. 
Taking into account the parity of the $\bar{s}$,  
the lowest energy state of $1/2^+$ requires the total 
orbital angular momentum to be odd and 
at least ${\bf L}=\sum_i {\bf l}_i={\bf 1}$, where the 
bold number denotes the angular momentum vector.
Since the $(ud)$ clusters are identical, we choose the Jacobi
coordinate to construct the spatial wavefunction. 
As a result~\cite{zhao-theta2}, the magnetic moment can be rewritten
as
\be
\mu_\theta\vsig=\frac{e_0}{6m_3}\vsig_3 +\frac{e_0}{6m_1}{\bf l}_\rho
+\frac{e_0}{6M_\Theta}\left(\frac{m_3}{m_1}+\frac{2m_1}{m_3}\right){\bf l}_\lambda \ ,
\ee
where ${\bf l}_\lambda$ and ${\bf l}_\rho$ are the orbital angular momenta
for the internal coordinates in the c.m. system, 
and ${\bf L}={\bf l}_\lambda+{\bf l}_\rho$;
$e_0$ is the unit charge. 
For the simplest case that either ${\bf l}_\lambda$ or ${\bf l}_\rho$
to be excited one unit, by taking the $z$-projection $\sigma_z=1$, we have
\be
\mu_\theta=\frac{e_0}{2M_\Theta}
\left[ -\frac{M_\Theta}{9m_3}+\frac{2M_\Theta}{9m_1} b^2 
+\frac{1}{9}\left(\frac{m_3}{m_1}+\frac{2m_1}{m_3}\right)a^2\right] ,
\ee
where $a^2$ and $b^2$ denote the probability 
of exciting ${\bf l}_\lambda$ and ${\bf l}_\rho$ by one unit, and $a^2+b^2=1$
as required by the normalization.
Adopting quark and diquark masses of Ref.~\cite{JW}:
$m_3=500$ MeV for $\bar{s}$ and $m_1=m_2=720$ MeV, we
have $\mu_\theta\simeq 0.094 (e_0/2M_\Theta)$ with $a^2=b^2=1/2$, 
$0.055 (e_0/2M_\Theta)$ with $a^2=1$ and $b^2=0$, 
and $0.133 (e_0/2M_\Theta)$ with $a^2=0$ and $b^2=1$. 
This rough estimate suggests a small magnetic moment for the $\Theta^+$
in the Jaffe and Wilczek's picture. We will take the value 
$\mu_\theta\simeq 0.13 (e_0/2M_\Theta)$, 
i.e. the anomalous magnetic moment $\kappa_\theta\simeq -0.87$,
in the calculation as follows.

\begin{figure}
\begin{center}
\epsfig{file=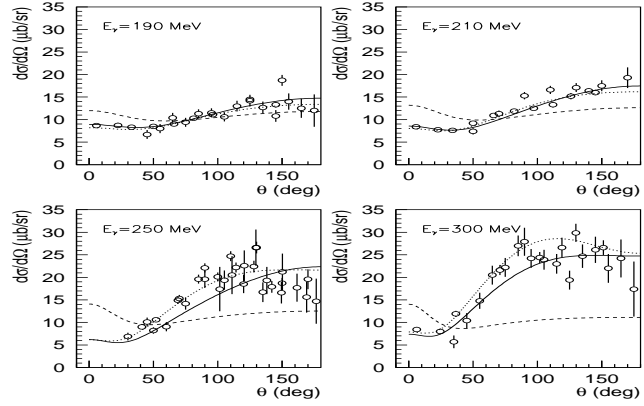, width=10cm,height=6.cm}
\caption{Differential cross sections for $\gamma n\to \pi^- p$
at $E_\gamma=190$, 210, 250, and 300 MeV. The solid curves
are results for Born terms plus Delta; dashed curves for 
Born terms; dotted curves for the full quark model 
calculations~\protect\cite{z-a-l-w}.
}
\protect\label{fig:(2)}
\end{center}
\end{figure}

To proceed, we make a nonrelativistic expansion of the above 
transition amplitudes, and adopt the typical nonrelativistic 
quark model wavefunctions for the form factor, which is essentially 
a Gaussian distribution for the wavefunction overlaps~\cite{z-a-l-w}.
This treatment is examined in $\gamma n\to \pi^- p$ at low energies. 
In Fig.~\ref{fig:(2)}, we present the calculation of cross sections
with the above effective Lagrangian. 
In comparison with the full quark model calculation 
(dashed curve)~\cite{z-a-l-w}, the form factor adopted account for 
the cross section reasonably well, and note that only the Delta resonance
is included here. 
This test also shows that the contact term and {\it t}-channel exchange
play important roles near threshold, which will be a guide
in the $\Theta^+$ production.

\begin{figure}
\begin{center}
\epsfig{file=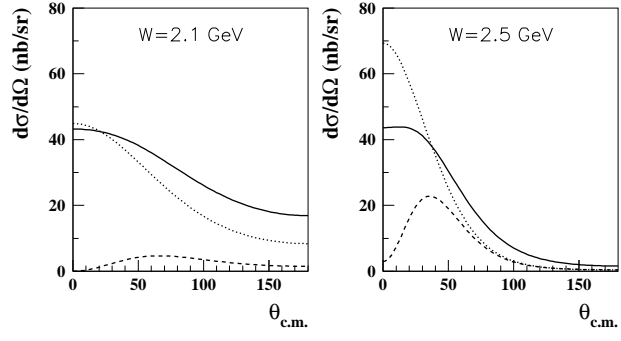, width=10cm,height=6.cm}
\caption{Differential cross sections for $\gamma n\to K^- \Theta^+$
with $\Theta^+$ of $1/2^+$ at $W=2.1$ and 2.5 GeV in the Born limit. 
The solid curves denote the results for the Born terms; the dashed curves
denote the results {\it without} the contact term; and the dotted curves
denote the results {\it without} the {\it s}- and {\it u}-channel
Born contributions. }
\protect\label{fig:(3)}
\end{center}
\end{figure}

In Fig.~\ref{fig:(3)}, the differential cross sections are presented
for $\gamma n\to K^-\Theta^+$ at $W=2.1$ and 2.5 GeV,
which exhibit forward peakings. 
The contact term accounts for the forward peaking and
is examined by suppressing it from contributing
as shown by the dashed curves. 
This feature is similar to the case of $\pi^-$ photoproduction.
We also examine the role played by the {\it s}- and {\it u}-channels
by switching off their contributions (dotted curves). It shows that 
the dominant contributions are from the contact term and {\it t}-channel
exchange. The bump structure arising from the dashed curves
are due to the relatively strong {\it t}-channel kaon exchange.
As we will see later, in contrast with the $\Theta^+$ of $1/2^-$,
the important role played by the contact term will be a signature 
for the $\Theta^+$ of $1/2^+$ in the polarized beam asymmetries
in the Born limit.

\begin{figure}
\begin{center}
\epsfig{file=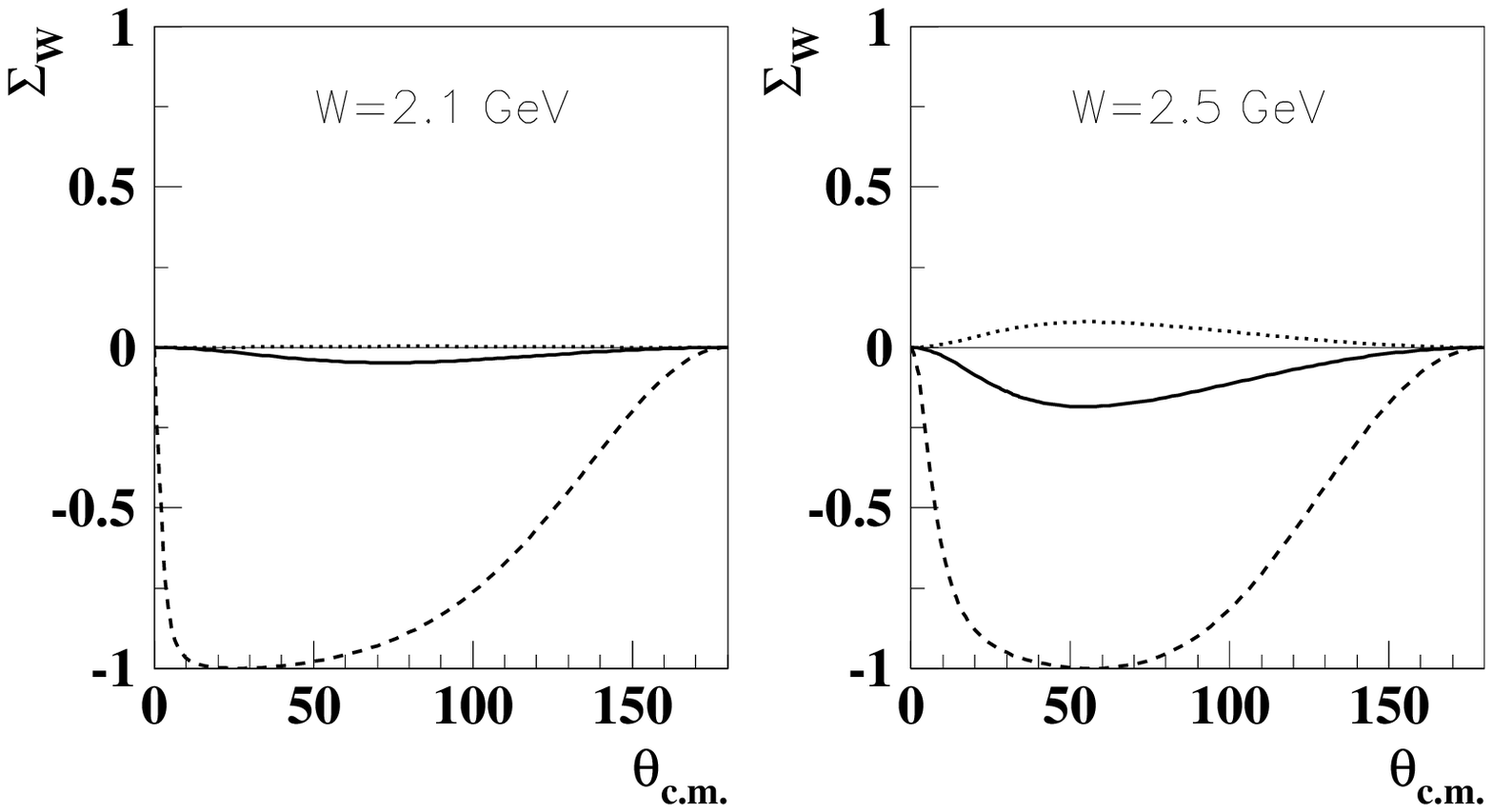, width=10cm,height=6.cm}
\caption{Polarized beam asymmetries for $\gamma n\to K^- \Theta^+$
with $\Theta^+$ of $1/2^+$
at $W=2.1$ and 2.5 GeV in the Born approximation limit.
The notations are the same as Fig.~\protect\ref{fig:(3)}.}
\protect\label{fig:(4)}
\end{center}
\end{figure}

Qualitatively, the photoproduction of $\Theta^+$ of $1/2^+$ 
is similar to the pseudoscalar meson photoproduction and will have 
the typical spin structures of the well-known CGLN amplitudes~\cite{CGLN}:
$\vsig\cdot\veps_\gamma$, 
$\vsig\cdot{\bf q}\vsig\cdot({\bf k}\times\veps_\gamma)$, 
$\vsig\cdot{\bf q}{\bf q}\cdot\veps_\gamma$, and 
$\vsig\cdot{\bf k}{\bf q}\cdot\veps_\gamma$. 
It makes the kinematics $\theta_{c.m.}=0^\circ$ and $180^\circ$
special since only the term of $\vsig\cdot\veps_\gamma$ will contribute.
This term will either raise or lower 
the initial neutron spin projection by one unit. Therefore, 
it will contribute to 
either $T_{-\frac 12,\lambda_\gamma \frac 12}$ 
or $T_{\frac 12,\lambda_\gamma -\frac 12}$
for a fixed polarization $\lambda_\gamma$. 
As a result, $\rho^1_{\frac 12,\frac 12}$
will vanish at $\theta_{c.m.}=0^\circ$ and $180^\circ$, and thus 
$\Sigma_W=0$ at these two scattering angles. 
More clear evidence for the importance of the contact term can be seen
in the polarized beam asymmetry through its interfences with other terms.
In Fig.~\ref{fig:(4)}, $\Sigma_W$ is presented
at two energies in parallel to Fig.~\ref{fig:(3)}. 
The asymmetries arising from the Born terms (solid curves) are relatively small,  
which are also similar to the case of $\pi^-$ photoproduction.
By switching off the contact term, we obtain the dashed curves, which 
are strongly stretched towards $-1$. This feature is accounted for 
by the dominant {\it t}-channel exchange (with the absence of the contact term)
in comparison with the relatively 
small {\it s}- and {\it u}-channels: the exclusive {\it t}-channel
will analytically lead to $\Sigma_W=-1$. 
However, due to the intereference of the contact transition, of which 
the exclusive contribution leads to vanishing asymmetries, 
the asymmetry exhibits a rather small deviations from zero. 
This feature is independent of the form factors since they will cancel
out. In this sense, additional information about the $\Theta^+$ 
can be derived from polarized beam asymmetries.
Nevertheless, the contact term in $1/2^+$ production
will provide a clear way to distinguish it from the $1/2^-$
if the transitions are only via the Born terms.

\subsection{Polarized beam asymmetry for $\Theta^+$ of $1/2^-$ in the Born limit}

For $\Theta^+$ of $1/2^-$, the effective Lagrangian for the $\Theta^+ n K$
system (Eq.~(\ref{lagrangian})) conserves parity and is gauge 
invariant. 
This suggests that the electromagnetic interaction will not contribute
to the contact term, which can be also seen in the leading 
term of nonrelativistic expansion, where the derivative operator
is absent. The Born approximation therefore includes three transitions
(Fig.~\ref{fig:(1)}(b), (c) and (d)):
{\it t}-channel kaon exchange, {\it s}-channel nucleon exchange,
 and {\it u}-channels $\Theta^+$ exchange.
The invariant amplitude can be written as
\be
{\cal M}_{fi}=M^t + M^s + M^u  \ ,
\ee
where the three transitions are given by
\bea
M^t & =&  -\frac{e_0g_{\Theta NK}}{t-M_K^2}\bar{\Theta}
(2q-k)_\mu A^\mu N, \nonumber\\ 
M^s & =&  g_{\Theta NK}\bar{\Theta} K
\frac{[\gamma\cdot(k + P_i) +M_n]}{s-M_n^2} 
\left[ e_n\gamma_\mu + \frac{i{\kappa}_n }{2M_n}\sigma_{\mu\nu} k^\nu\right]
A^\mu N,
\nonumber\\
M^u & =&  g_{\Theta NK}\bar{\Theta} 
\left[ e_\theta\gamma_\mu + \frac{i{\kappa}_\theta }{2M_\Theta}
\sigma_{\mu\nu} k^\nu\right] A^\mu 
\frac{[\gamma\cdot(P_f-k)+M_\Theta]}{u-M_\Theta^2} 
 N K, 
\eea
where the coupling constant 
$g_{\Theta NK}=[4\pi M_\Theta \Gamma_{\Theta^+\to K^+ n}
/|{\bf p}^\prime|(E_n+M_n)]^{1/2}$ 
has a range of 
0.43$\sim$ 0.96 with $\Gamma_{\Theta^+\to K^+ n}=5$ to 25 MeV, and turns out to
be much smaller than $1/2^+$ coupling. We adopt $g_{\Theta NK}=0.61$
corresponding to $\Gamma_{\Theta^+\to K^+ n}=10$ MeV in the calculation.

If $\Theta^+$ has spin-parity $1/2^-$, one may simply estimate its
magnetic moment as the sum of $(u\bar{s})$ and $(udd)$ clusters, 
of which the relative orbital angular momentum is zero. 
Assuming these two clusters are both color singlet, the 
total magnetic moment of this system can be written as
\be
\mu_\theta =\left(\frac{2e_0}{6m_u}+\frac{e_0}{6m_s}\right)
-\frac{e_0}{3m_u}=\frac{e_0}{6m_s},
\ee
which leads to a small anomalous magnetic moment taking into account
$3m_s\simeq M_\Theta$.
However, since such a simple picture may not be sufficient, we 
also include $\kappa_\theta=\pm\kappa_p=\pm 1.79$ to make a sensitivity
test.

\begin{figure}
\begin{center}
\epsfig{file=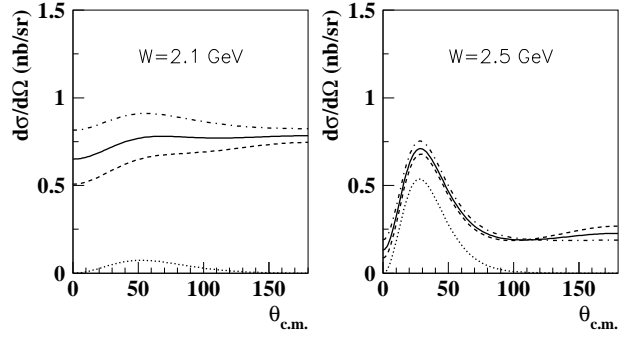, width=10cm,height=6.cm}
\caption{Differential cross sections for $\gamma n\to K^- \Theta^+$
with $\Theta^+$ of $1/2^-$ at $W=2.1$ and 2.5 GeV 
in the Born limit.
The solid curves denote the results for the Born terms; 
the dotted curves denote the results for the exclusive {\it t}-channel; 
the dashed and dot-dashed curves denote the results for 
$\kappa_\theta=\pm 1.79$, respectively. }
\protect\label{fig:(5)}
\end{center}
\end{figure}

In Fig.~\ref{fig:(5)}, the differential cross sections for $1/2^-$
production are presented at $W=2.1$ and 2.5 GeV. 
The solid curves are results of the Born terms, while the 
dotted curves denote calculations for the exclusive {\it t}-channel, i.e.
with the {\it s}- and {\it u}-channel Born terms eliminated. 
Near threshold, the {\it s}- and {\it u}-channels turn out 
to be important as shown by the dotted curve, and explains
the rather flat distribution. Above threshold, 
the {\it t}-channel kaon exchange becomes more and more important 
and produces a bump structure in the forward angle region. 
The strong forward peaking appearing in $1/2^+$ is absent here.
The cross sections for $\kappa_\theta=\pm 1.79$ (dashed and dot-dashed)
are also presented. In comparison with the solid curves, 
quite significant sensitivity to  $\kappa_\theta$ appears near threshold.

\begin{figure}
\begin{center}
\epsfig{file=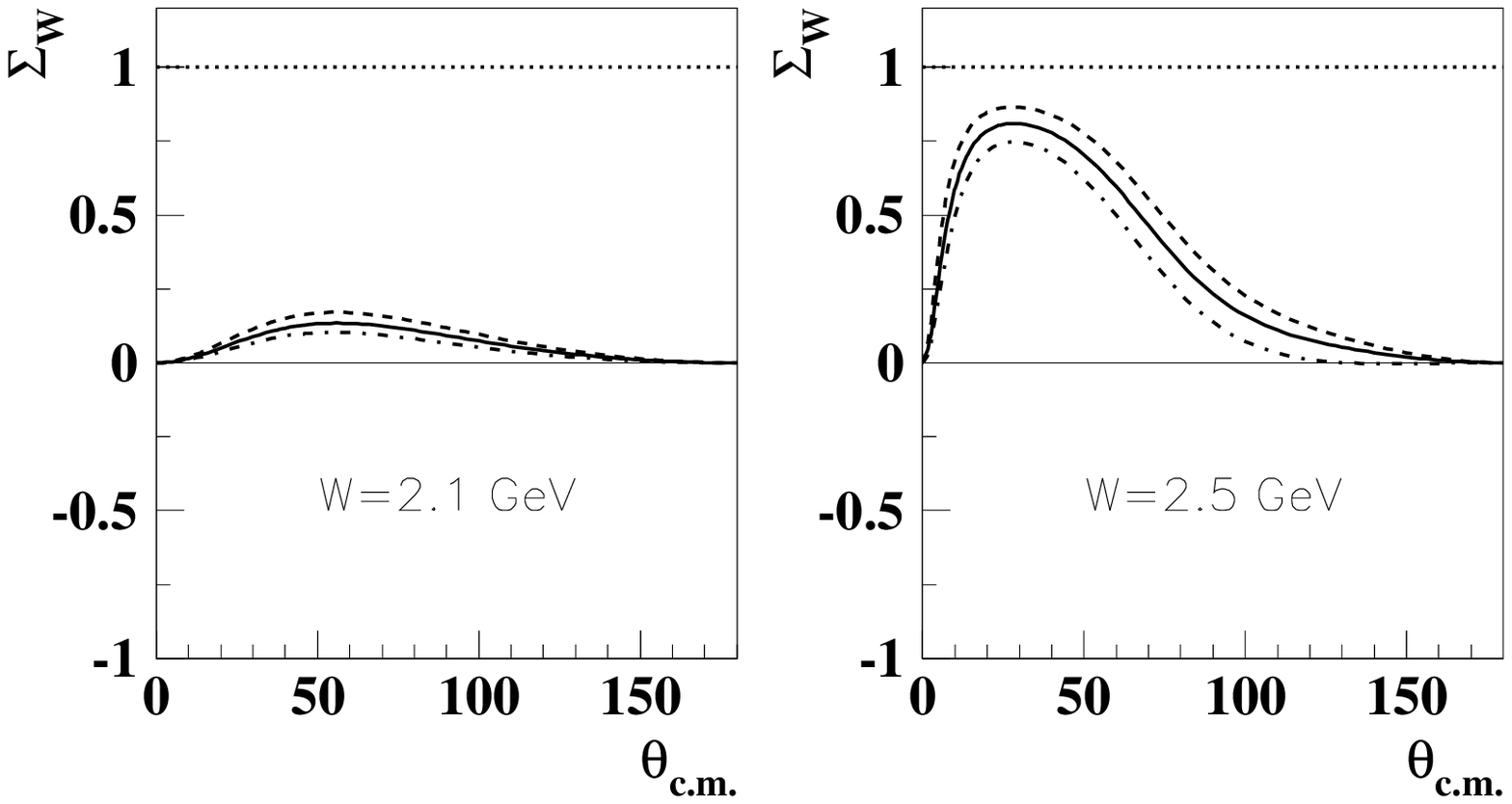, width=10cm,height=6.cm}
\caption{Polarized beam asymmetries for $\gamma n\to K^- \Theta^+$
with $\Theta^+$ of negative parity at $W=2.1$ and 2.5 GeV 
in the Born approximation limit.
The notations are the same as Fig.~\protect\ref{fig:(5)}.}
\protect\label{fig:(6)}
\end{center}
\end{figure}

The {\it t}-channel dominance above the threshold region
leads to specific signature of the $1/2^-$ in the polarization asymmetries.
Qualitatively, 
given an exclusive contribution from the 
{\it t}-channel kaon exchange, the transition amplitude will have a 
structure of ${\bf q}\cdot\veps_\gamma$, which is spin-independent.
In this circumstance, the polarized beam asymmetry becomes
$\Sigma_A=-1$ (or $\Sigma_W=+1$) 
due to the exact cancellation of the angular parts 
in $\rho^0_{\frac 12,\frac 12}$ and $\rho^1_{\frac 12,\frac 12}$. 
In Fig.~\ref{fig:(6)}, the asymmetry $\Sigma_W$ 
produced by the Born terms is shown by the solid curves at 
$W=2.1$ and 2.5 GeV. The dotted lines denote the constant asymmetry $+1$
produced by the exclusive {\it t}-channel kaon exchange.
The inclusion of the {\it s}- and {\it u}-channels violates the constant
$+1$ asymmetry, and 
results in the vanishing asymmetries at $\theta_{c.m.}=0^\circ$ and $180^\circ$.
In particular, due to the dominance of the {\it s}- and {\it u}-channels
near threshold, the asymmetry appears to be small. With the inceasing 
energies the {\it t}-channel becomes
more and more important, and will stretch the asymmetries ($\Sigma_W$) 
towards $+1$ with large values.
This is characteristic for $1/2^-$, and makes it 
different from the moderate behavior of $1/2^+$, e.g., at $W=2.5$ GeV. 
We also present the results for 
$\kappa_\theta=\pm 1.79$ (dashed and dot-dashed). 
Quite sizeable effects appear at middle angles, 
while the vanishing asymmetries
at forward and backward angles are not sensitive to 
the $\Theta^+$'s anomalous magnetic moment.

Certain points can be learned here:

i) For the Born terms, the $\Theta^+$ of different parities
results in an order of magnitude difference in the cross sections. 
This is consistent with other model calculations~\cite{nam,liu-ko,oh}. However, 
it also shows that the cross section is very sensitive to the
empirical form factors adopted, which may lead to large 
uncertainties in the model predictions. 

ii) The polarized beam asymmetry has less dependence 
on the form factor than the cross section. 
For the Born terms, as shown by Figs.~\ref{fig:(4)} and \ref{fig:(6)}, 
the asymmetries at $W=2.5$ GeV exhibit completely different
behaviors for the $\Theta^+$ of $1/2^+$ and $1/2^-$, 
which should make these two configurations distinguishable. 

iii) For exclusive contributions, e.g. the contact term plus {\it t}-channel 
in the production of $1/2^+$, and the {\it t}-channel of $1/2^-$, 
we also find significantly different features arising from 
the polarized beam asymmetries. 

Summarily, supposing the $\Theta^+$ is produced 
dominantly via Born terms,  the polarized beam 
asymmetry measurement incorporating with the cross section
will be able to determine its quantum numbers.
However, in reality such an assumption is too 
drastic since other mechanisms may interfer and even play 
dominant roles with increasing energies.
To proceed and investigate the effects from other mechanisms, 
we will include the vector meson $K^*$ exchange in the model and 
study its impact on the cross sections and polarized beam asymmetries.

\subsection{$K^*$ exchange in $\Theta^+$ of $1/2^+$ production}

The role played by vector meson exchanges 
in pseudoscalar meson photoproduction 
is still a controversial issue in theoretical phenomenology. 
According to the duality argument~\cite{dolen}, 
the inclusion of vector meson exchange in association with 
a complete set of {\it s}-channel resonances may result in 
double-counting of the cross sections. Therefore, one in principle, 
can include either all {\it s}-channel states or a complete set of 
{\it t}-channel trajectories in a full calculation. 
In reality, difficulty arises from the lack of knowledge about 
these two extremes, and empirical phenomenologies
will generally include the leading contributions 
of both channels taking into account partially their 
different roles at different kinematical regions.

The exotic $\Theta^+$ production would complicate such an issue
if any {\it s}- and {\it t}-channel processes beyond the Born approximation
are important. 
Certainly, this relevance deserves a full independent study. 
In this work, we will only empirically include the 
$K^*$ meson exchange in $\gamma n \to K^- \Theta^+$. 
The essential concern here is its effects on the polarized beam 
asymmetries: will it change  the basic features 
produced by the Born terms near threshold?

The effective Lagrangian for $K^*K\gamma$ is given by 
\be
{\cal L}_{K^*K\gamma}=\frac{ie_0g_{K^*K\gamma}}{M_K}
\epsilon_{\alpha\beta\gamma\delta}
\partial^\alpha A^\beta\partial^\gamma V^\delta K  + \mbox{H.c.} \ ,
\ee
where $V^\delta$ denotes the $K^*$ field; $g_{K^*K\gamma}=0.744$ is  
determined by the $K^{*\pm}$ decay width 
$\Gamma_{K^{*\pm}\to K^\pm\gamma}=50$ keV~\cite{pdg2000}.

The $K^*N\Theta$ interaction is given by 
\be 
{\cal L}_{\Theta N K^*}=g_{\Theta N K^*}\bar{\Theta}
(\gamma_\mu +\frac{\kappa_\theta^*}{2M_\Theta}
\sigma_{\mu\nu}\partial^\nu)V^\mu N   + \mbox{H.c.} \ ,
\ee
where $g_{\Theta N K^*}$ and $\kappa_\theta^*$ denote
the vector and tensor couplings, respectively.
So far, there is no experimental information about these two 
couplings. A reasonable assumption based on 
an analogue to vector meson exchange in pseudoscalar meson production
is that $|g_{\Theta N K^*}|=|g_{\Theta N K}|$.
For the tensor coupling, we assume $|\kappa_\theta^*|=|\kappa_\rho|= 3.71$, 
the same as $\rho NN$ tensor coupling but with an arbitary phase.
Therefore, four sets of different phases are possible.

\begin{figure}
\begin{center}
\epsfig{file=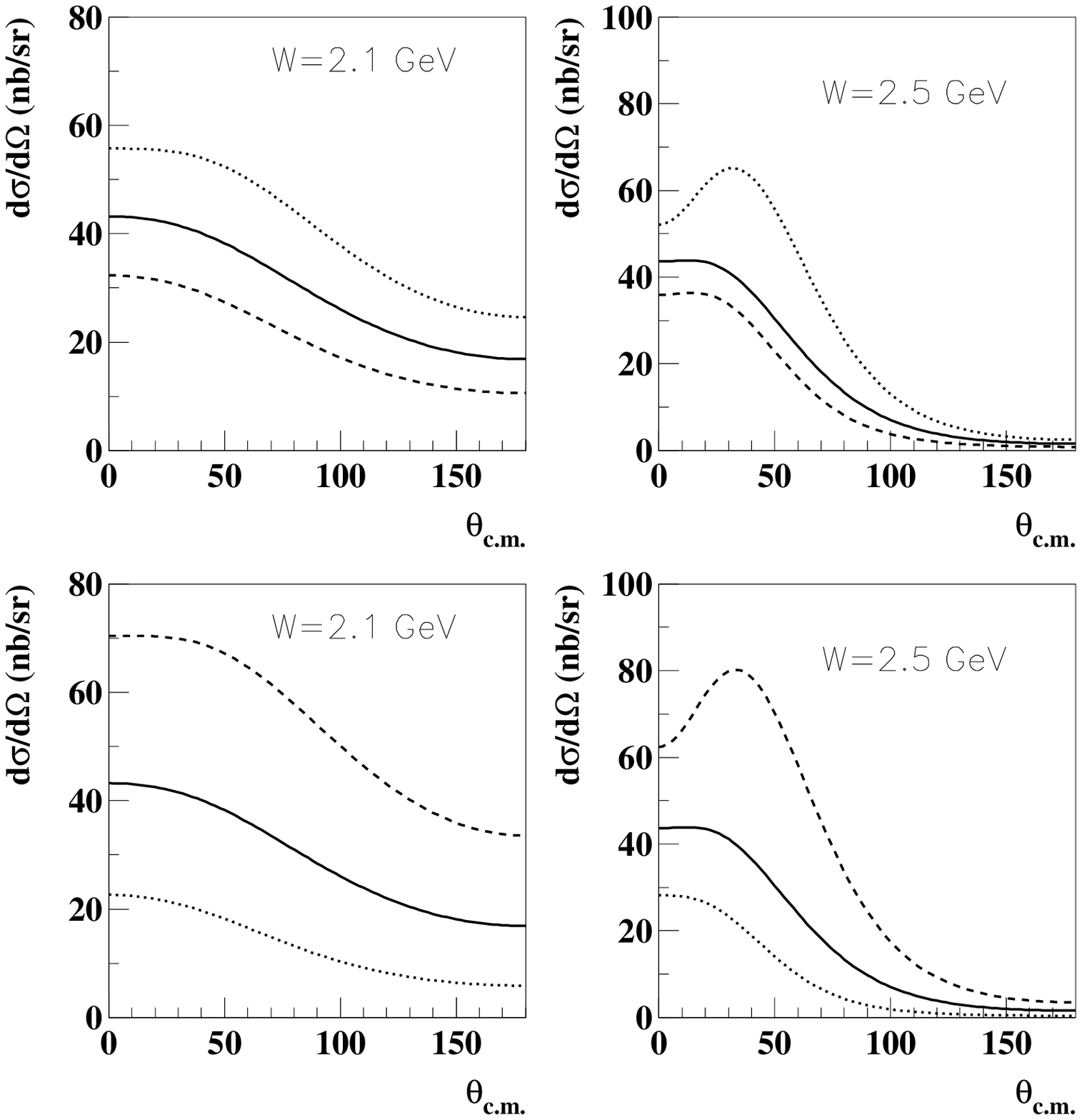, width=10cm,height=10.cm}
\caption{Differential cross sections for 
$\Theta^+$ of $1/2^+$ at $W=2.1$ and 2.5 GeV with the $K^*$ exchange. 
The solid curves are results in the Born limit, while the dashed and
dotted curves denote results with the $K^*$ exchange included 
with different phases: $(g_{\Theta N K^*}, \kappa_\theta^*)= (-2.8,-3.71)$ 
(dashed curves in the upper two figures),
$(+2.8,+3.71)$ (dotted curves in the upper two figures), 
$(-2.8, +3.71)$ (dashed curves in the lower two figures), and 
$(+2.8, -3.71)$ (dotted curves in the lower two figures).}
\protect\label{fig:(7)}
\end{center}
\end{figure}

\begin{figure}
\begin{center}
\epsfig{file=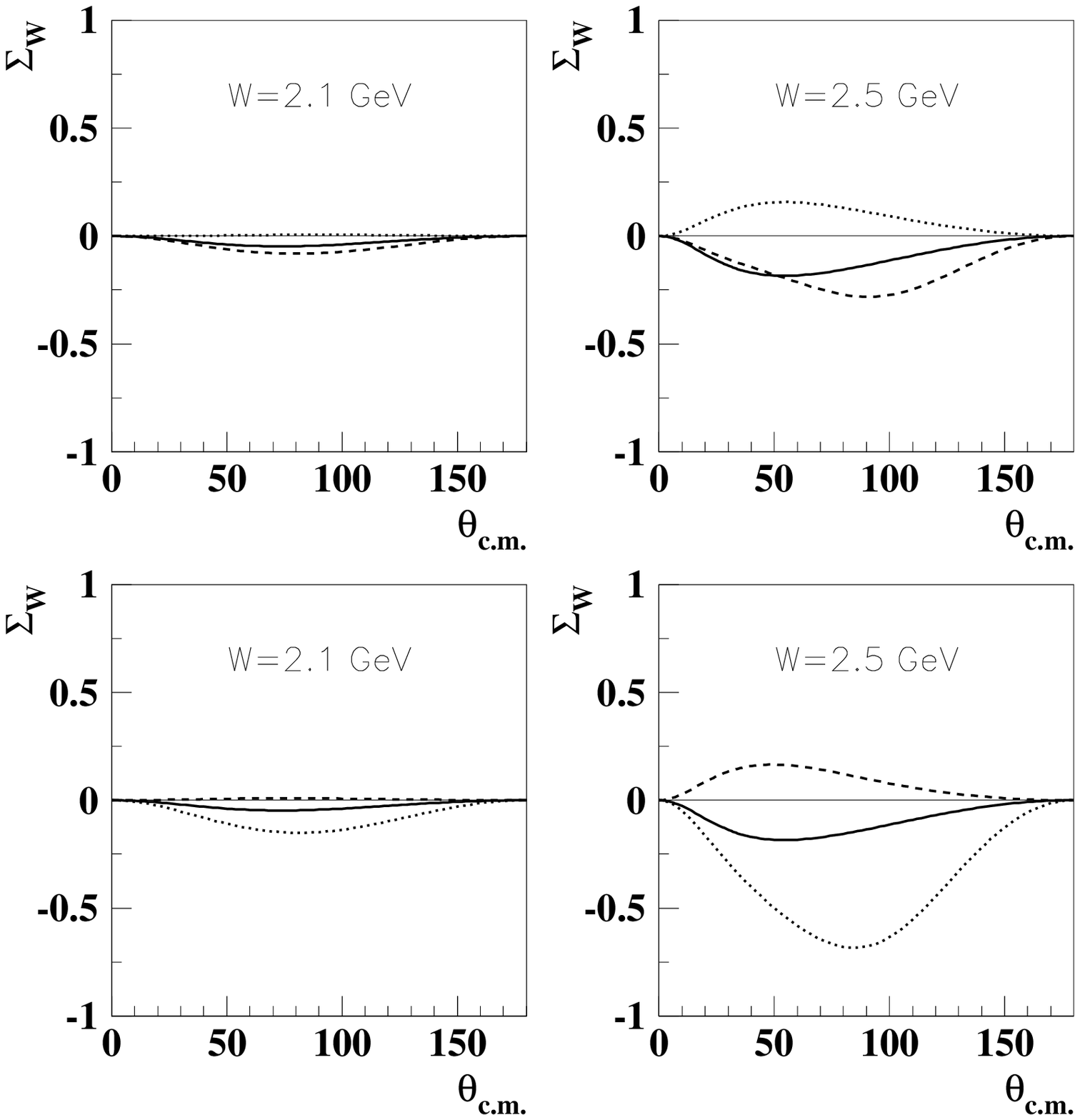, width=10cm,height=10.cm}
\caption{Polarized beam asymmetries for 
$\Theta^+$ of $1/2^+$ at $W=2.1$ and 2.5 GeV with the $K^*$ exchange.
The notations are the same as Fig.~\protect\ref{fig:(7)}.}
\protect\label{fig:(8)}
\end{center}
\end{figure}

In Fig.~\ref{fig:(7)}, the differential cross sections for 
$1/2^+$ production at $W=2.1$ and 2.5 GeV are presented. 
The four possible phases: 
$(g_{\Theta N K^*}, \kappa_\theta^*)= (-2.8,-3.71)$ 
and $(+2.8, +3.71)$ corresponding to the dashed  and dotted curves
in the upper two figures, and $(-2.8, +3.71)$ and $(+2.8,-3.71)$
to the dashed  and dotted curves in the lower two figures,
are shown in contrast with
the calculations without the $K^*$ exchange (solid curves).
It shows that the $K^*$ exchange with 
its coupling comparable 
with $g_{\Theta NK}$ will play an important role. 
The change of the parameter phases results in significant changes
to the cross sections. In particular, the tensor coupling 
turns out to be dominant.
The interference of the $K^*$ exchange in 
the polarized beam asymmetry also becomes important. 
In Fig.~\ref{fig:(8)}, the polarized beam asymmetry is presented
in parallel to Fig.~\ref{fig:(7)}. It shows that near threshold
the effects from the $K^*$ exchange is quite small. 
In contrast with its effects in the differential cross section 
near threshold, the small change to the asymmetries
can be understood by the overall enhancement or cancellation 
within the amplitudes throughout the scattering angles as shown 
by the curves for $W=2.1$ GeV in Fig.~\ref{fig:(7)}.
At $W=2.5$ GeV quite significant asymmetries are produced 
by the phase changes.

\subsection{$K^*$ exchange in $\Theta^+$ of $1/2^-$ production}

\begin{figure}
\begin{center}
\epsfig{file=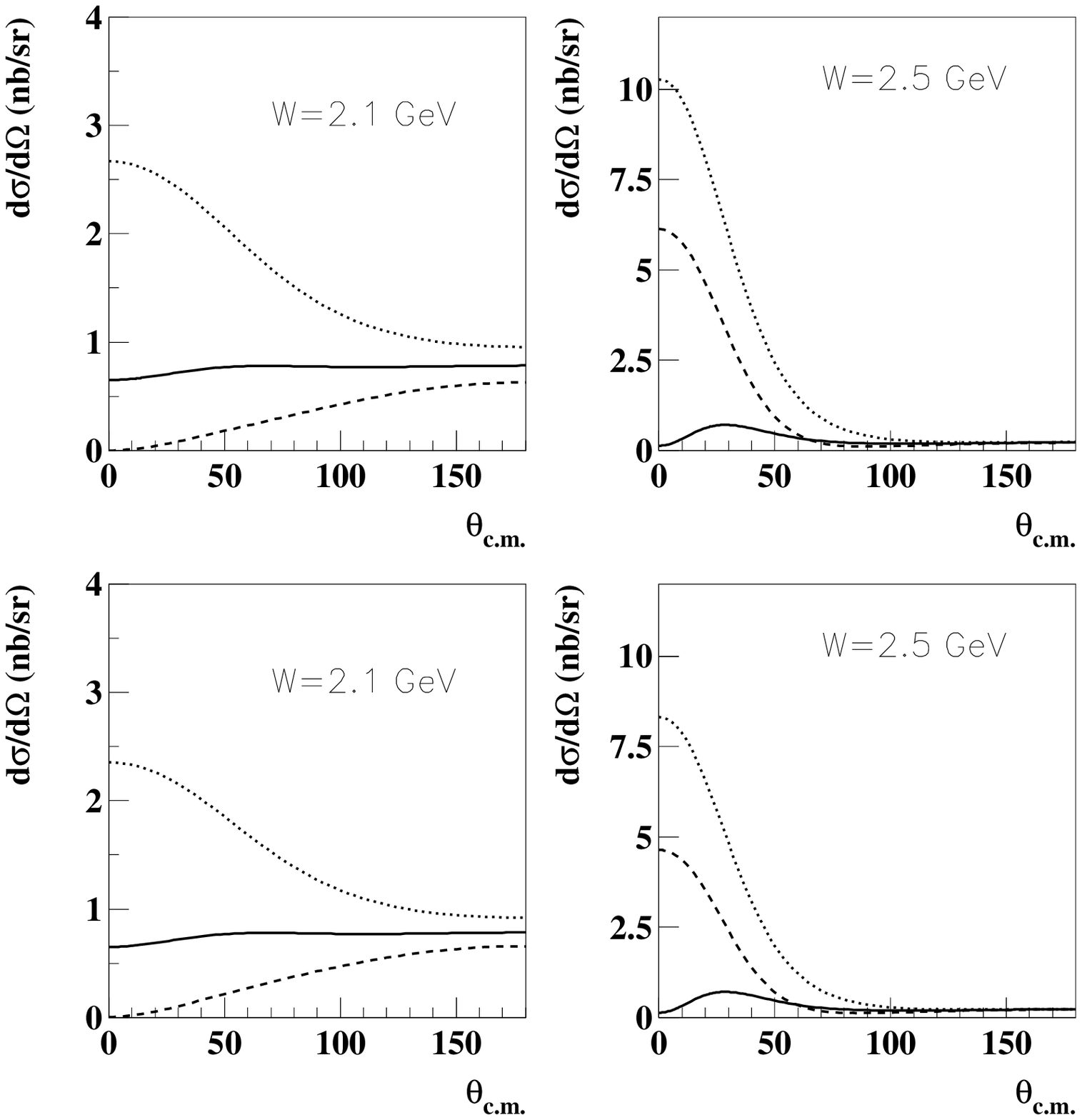, width=10cm,height=10.cm}
\caption{Differential cross sections for 
$\Theta^+$ of $1/2^-$ at $W=2.1$ and 2.5 GeV with the $K^*$ exchange. 
The solid curves are results in the Born limit, while the dashed and
dotted curves denote results with the $K^*$ exchange included 
with different phases: $(g_{\Theta N K^*}, \kappa_\theta^*)= (-0.61,-0.371)$ 
(dashed curves in the upper two figures),
$(+0.61,+0.371)$ (dotted curves in the upper two figures), 
$(-0.61, +0.371)$ (dashed curves in the lower two figures), and 
$(+0.61, -0.371)$ (dotted curves in the lower two figures).}
\protect\label{fig:(9)}
\end{center}
\end{figure}

We also include the $K^*$ exchange in the $1/2^-$ production, in which
the $K^*N\Theta$ interaction is given by 
\be 
{\cal L}_{\Theta N K^*}=g_{\Theta N K^*}\bar{\Theta}
\gamma_5(\gamma_\mu +\frac{\kappa_\theta^*}{2M_\Theta}
\sigma_{\mu\nu}\partial^\nu)V^\mu N   + \mbox{H.c.} \ .
\ee
Similar to the case of $1/2^+$, we assume
$|g_{\Theta N K^*}|=|g_{\Theta N K}|=0.61$. 
In the calculation we choose the anomalous magnetic moment
$|\kappa_\theta^*|=0.371$, which is ten times smaller than 
that in the production of $1/2^+$. This choice is based on 
that $g_{\Theta N K}=0.61$ is also about one magnitude smaller 
than that in the production of $1/2^+$.
In Fig.~\ref{fig:(9)}, the cross section for $1/2^-$ production
at $W=2.1$ and 2.5 GeV are presented.
The four possible phases:
$(g_{\Theta N K^*}, \kappa_\theta^*)= (-0.61,-0.371)$ and $(+0.61, +0.371)$
corresponding to the dashed and dotted curves in the upper two figures, 
and $(-0.61, +0.371)$ and $(+0.61,-0.371)$ to the dashed and dotted 
curves in the lower two figures,
are shown in contrast with
the calculations without the $K^*$ exchange (solid curves).
It shows that the forward cross section is dominated by the
$K^*$ exchange. Near theshold, the cancellation 
between the Born terms and $K^*$ exchange of two phases can even 
lead to vanishing cross sections, while the other two phases
lead to forward peaking.
Above the threshold region, 
the Born terms become relatively small and forward peaking 
due to the $K^*$ exchange appears in those four sets of phases.

A rather interesting feature arising from the $K^*$ exchange in 
the production of $1/2^-$
is that the cross sections increase drastically. 
For one of the phases, $(g_{\Theta N K^*}, \kappa_\theta^*)= 
(+0.61,+0.371)$ at $W=2.5$ GeV (dotted curve in the upper-right figure of 
Fig.~\ref{fig:(9)}), 
the cross sections at forward angles are comparable 
with the $1/2^+$ production with 
$(g_{\Theta N K^*}, \kappa_\theta^*)=(+2.8, -3.71)$ 
(dotted curve in the lower-right figure 
of Fig.~\ref{fig:(7)}).
As mentioned at the beginning 
that so far little information about the form factors 
has been available. Therefore, large uncertainties could be with the 
model predictions for the cross sections. 
In this sense, a single measurement of the cross sections 
may not be able to distinguish these two parities if the 
$K^*$ exchange has large contributions.

\begin{figure}
\begin{center}
\epsfig{file=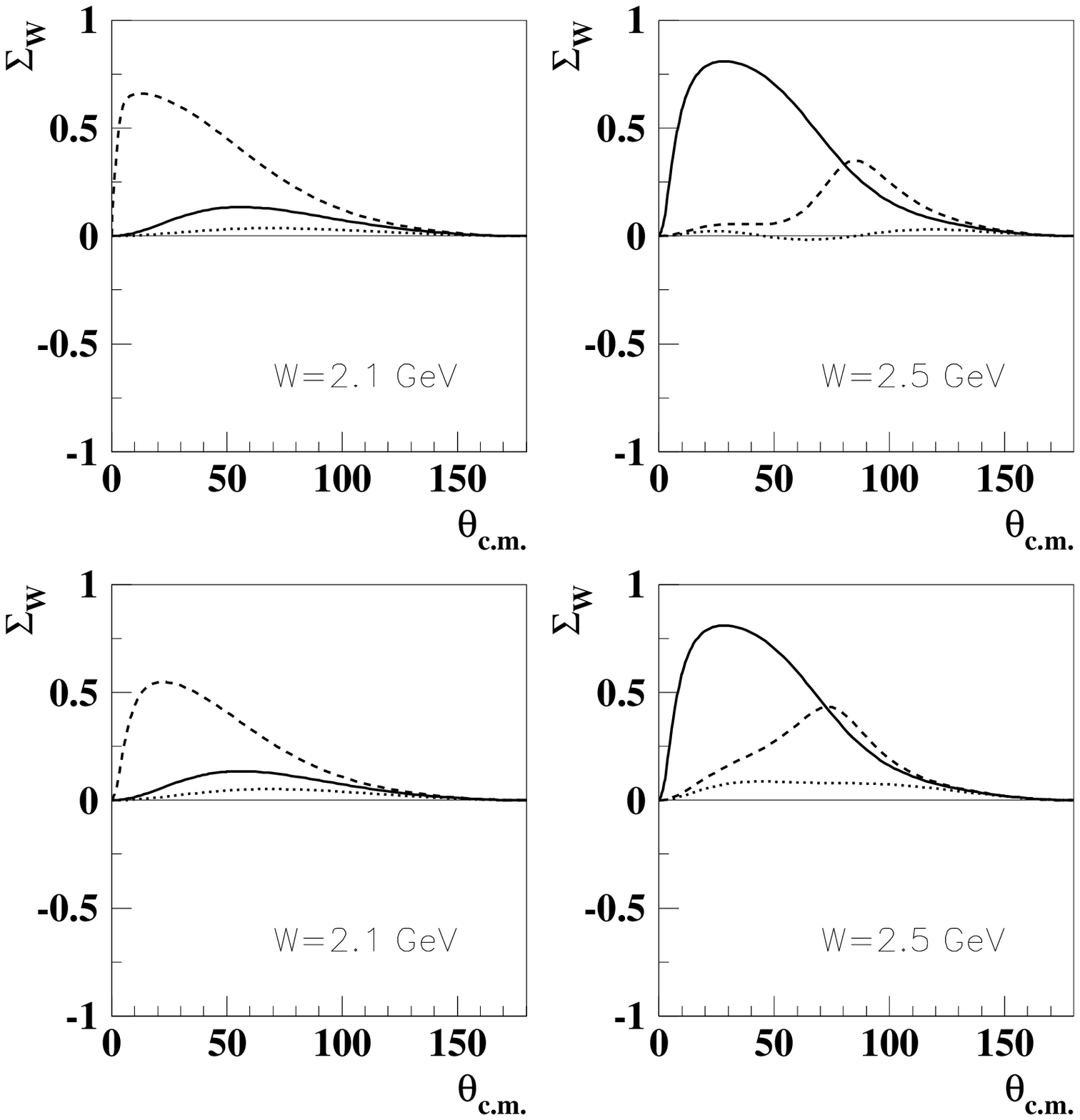, width=10cm,height=10.cm}
\caption{Polarized beam asymmetries for 
$\Theta^+$ of $1/2^-$ at $W=2.1$ and 2.5 GeV with the $K^*$ exchange.
The notations are the same as Fig.~\protect\ref{fig:(9)}.}
\protect\label{fig:(10)}
\end{center}
\end{figure}

More information about the $\Theta^+$ properties can be obtained 
from the polarization asymmetries. 
In Fig.~\ref{fig:(10)}, we present the calculations for $\Sigma_W$ 
in parallel to the cross sections.
Significant sensitivities to the $K^*$ exchange are found, and 
interesting feature arises from its interferences. 
For $(g_{\Theta N K^*}, \kappa_\theta^*)=(-0.61,\mp 0.371)$ 
(dashed curves in the upper and lower figures), 
significantly large positive asymmetries appear near threshold. 
This is a unique feature since large
positive asymmetries cannot be produced in all other 
phase sets and in the production of $1/2^+$. 
However, as shown by the dotted curves, asymmetries produced 
by phase sets $(+0.61,\pm 0.371)$ are generally small.
These could mix up with the production of $1/2^+$ near threshold
if the measurement is only carried out in that energy region.
Fortunately, taking into account the energy evolution of the asymmetries 
and cross sections, these two parities are still distinguishable 
from each other.

\section{Discussions and summaries}

We investigated the possibility of using 
polarized photon beams to 
distinguish the quantum numbers of $\Theta^+$ in experiment. 
The polarization of the photon beams 
transfers information of the $\Theta^+$ configurations
through the interference among the transition amplitudes, 
from which insights into the dynamical role played by the 
$\Theta^+$ can be gained.
We first examined the kinematics by comparing 
the angular distributions of $\Theta^+$ decays 
with different quantum numbers.
It showed that a clear $\cos^2\theta$ distribution after the sum over 
all the forward production events of all $\phi$-angles would 
be evidence for $3/2^+$ configuration.  
For $\Theta^+$ of $1/2^+$ and $1/2^-$, 
the difference arising from the angular distributions of $\Theta^+$ decay 
is strongly related to the dynamics of the production mechanism. 
In the Born approximation limit, the decay distributions
for these two configurations will both become isotropic. 
Because of this, a coherent investigation of the 
production dynamics and the kinematical analysis is necessary.

So far, the theoretical calculation of the cross sections 
has been strongly model-dependent though it suggests
significant difference between $1/2^+$ and $1/2^-$. 
For instance, the coupling strength $g_{\Theta NK}$ is determined 
by the $\Theta^+$ decay width, which however has still large uncertainties.
Also, the form factors as well as the roles played by other mechanisms, 
e.g., $K^*$ exchange and spin 3/2 state,
are almost unknown. Due to such complexities,
the measurement of cross sections 
could be a premier reference for distinguishing these two 
configurations, but hardly the only one.
Other observables are also needed not only to pin down 
the $\Theta^+$'s quantum numbers, but also 
constrain theoretical phenomenologies.

Based on the investigations of this work, 
the following points concerning  
the $\Theta^+$'s quantum numbers and the photoproduction mechanisms 
can be learned:

i) It seems very likely that a flat distribution near threshold
will unambiguously refer to $1/2^-$ for $\Theta^+$. For other 
situations, measurements of multi-observables are needed.

ii) With the same total width for the $\Theta^+$, and assuming
the same form factors, the 
cross section for $1/2^+$ is much larger than  $1/2^-$ in the Born limit.
This is consistent with other model calculations~\cite{nam,liu-ko,oh}.

iii) The exclusive Born terms of $1/2^+$ production generally produce
small negative values of asymmetries, while those of $1/2^-$ produce
rather large positive asymmetries above the threshold region. 
This feature is useful if the Born terms are the dominant contributions
in the $\Theta^+$ production.

iv) The polarized beam asymmetry turns out to be sensitive 
to the $K^*$ exchange, which eventually requires a better understanding of
the $\Theta N K^*$ coupling.
In the case of weak $\Theta N K^*$ coupling, the measurement of 
polarized beam asymmetries above 
the threshold region will be decisive on the $\Theta^+$'s quantum numbers
as summarized in iii).
For a strong $\Theta N K^*$ coupling as $\Theta N K$, 
although the situation will be complicated,  
the asymmetry measurement in association with the cross section 
can still provide more information about the $\Theta^+$.

v) Contributions from other spin states in the {\it s}- and {\it u}-channels 
may also become important. As shown by the dotted curves in Fig.~\ref{fig:(3)}, 
the {\it s}- and {\it u}-channel Born terms have sizeable  contributions
even above the reaction threshold. Other states, e.g., spin 3/2 one, 
may produce non-negligible 
effects at certain kinematics.  

In brief, incorporating with the cross section observable, 
it seems likely that the polarized beam asymmetry
will be able to provide additional information about the $\Theta^+$ 
properties, which should be useful for determining its quantum numbers.
The full angle detectors at SPring-8, JLab, ELSA, and ESRF
will has a great advantage for such an exploration.

\section*{Acknowledgement}

Special acknowledgement goes to F.E. Close for generously 
sharing many of his enlighting ideas and suggestions, 
and for many instructive comments on this work. 
The author thanks J.S. Al-Khalili for many useful discussions 
during preparing this paper. 
The author also thanks K. Hicks and W.-J. Schwille for responses 
related to their experiments. 
Useful communications with S.L. Zhu, and J. Kellie are acknowledged.
This work is supported
by grants from 
the U.K. Engineering and Physical
Sciences Research Council (Grant No. GR/R78633/01).


\end{document}